\documentclass{iopart}
\usepackage{iopams}
\usepackage{graphicx}
\newcommand{\abs}[1]{\left\vert #1 \right\vert}

\newcommand{\tH}{\widetilde{H}}
\newcommand{\trho}{\widetilde{\brho}}

\begin{document}

\title{Edge Induced Qubit Polarization in Systems with Ising Anyons}

\author{David J.~Clarke and Kirill Shtengel}

\address{Department of Physics and Astronomy, University of California,
Riverside, CA 92521, USA}
\begin{abstract}
We investigate the proposed experimental setup for measuring the
topological charge in a an Ising anyon system by means of Fabry-P\'{e}rot interferometry with a chiral edge state. We show that such an interferometer has the unintended but not necessarily
unwelcome effect of stabilizing the state of the system being measured (i.e., a
topological qubit). We show further that interactions between the edge mode and
the localized bulk quasiparticles can have the effect of polarizing the qubit,
stabilizing its state. We discuss the these results in the context of recent interferometer experiments in the $\nu=5/2$ fractional quantum Hall state, where the first of these effects is small, but the second may be relevant to the observed phenomena.
\end{abstract}

\submitto{\NJP}

\section{Introduction} 
\begin{figure}[htb]
\begin{center}
\includegraphics[width=0.6\columnwidth]{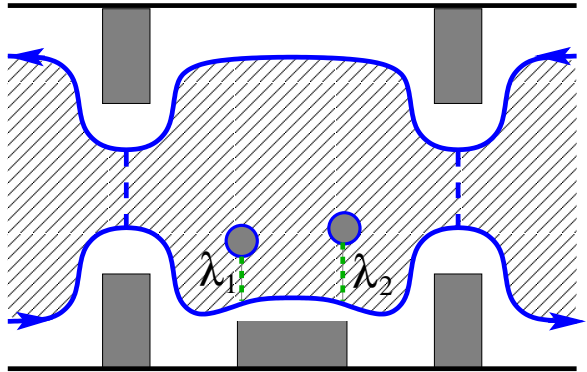}
\caption{A typical
Fabry-P\'{e}rot quasiparticle interferometer in the fractional quantum Hall
setting. The interference  signal results from two alternative paths taken by
the probe quasiparticles entering the interferometer along the lower edge,
tunnelling via one of the two quantum point contacts and leaving along the upper
edge. The interference pattern depends on the fusion channel (overall
topological charge) of the two localized non-Abelian anyons forming a qubit
between the quantum point contacts, which in turn is affected by the coupling
between these localized anyons and the edge.}
\label{fig:interferometer}
\end{center}
\end{figure}

Non-Abelian anyons are expected to occur in a number of
condensed matter systems, with the most prominent example being the $\nu=5/2$
fractional quantum Hall (FQH) state \cite{Moore91,Greiter92} where some
experimental evidence supports their existence \cite{Willett09a,Willett10a}.
More recently, a slew of systems with topological superconductivity (either
intrinsic or induced by a proximity effect on the boundary of a topological
insulator or a semiconductor with strong spin-orbit coupling) that may
support non-Abelian excitations have been proposed as well
\cite{Fu08,Sau10a,Alicea10a,Lee09}. Should non-Abelian statistics indeed be
confirmed experimentally, then quantum information could potentially be stored
in the combined state (fusion channel) of these quasiparticles and manipulated
in a non-local, and therefore fault-tolerant, fashion
\cite{Kitaev03,Preskill98,Freedman03b}. Current proposals for measuring such
quantum information involve quasiparticle
interferometery\cite{Fradkin98,DasSarma05,Bonderson06b,Bonderson08a}. In a
typical quantum Hall setting, the ``arms'' of a Fabry-P\'{e}rot interferometer
are formed by the FQH edges, with two quantum point contacts acting as beam
splitters as shown in \fref{fig:interferometer}. (Similar proposals were also
adopted in the context of heterostructures of superconducting and either
topologically insulating or semiconducting layers
\cite{Akhmerov09a,Fu09b,Nilsson10a,Sau10c}.) Such a measurement scheme relies in
a crucial way on the assumption that the topological charge of quasiparticles
inside the interferometric loop does not change during the time required to
measure it. This leads to a common concern about the feasibility of this scheme
stemming from the fact that interactions between bulk quasiparticles and the
edge modes have the potential of rapidly changing the state of the bulk, which
in turn will ``wash out'' the expected interference pattern.  This makes one
question even the utility of such interferometers for probing the non-Abelian
nature of quasiparticle statistics, much less their reliability for storing
quantum information. In other words, the question arises of how \emph{any}
interference signal might be seen in experiments like those reported by Willett
et al.\cite{Willett09a,Willett10a} in spite of bulk-edge interactions that are
estimated to be significant.

It has been shown that strong quasiparticle tunnelling leads to effective
absorbtion of quasiparticles by the edge \cite{Rosenow08a,Bishara09b,Rosenow09a}.
However, the presence of multiple quasiparticles in the interferometric loop
means that generally, some quasiparticle will be far enough away from the edge
to remain unabsorbed, yet have an appreciable coupling to the edge. This would lead to
the rapidly fluctuating topological charge inside the loop. The purpose of this note is to describe two additional effects that may, in principle,
counter the effect of such fluctuation and lead to an observable interference
signal in the presence of bulk-edge coupling.

First, a continuous measurement of the aforementioned type may have the
secondary effect of stabilizing the state of the qubit being measured via the
so-called ``quantum Zeno effect'' \cite{Misra77}. In Section~\ref{zeno}, we
describe the stabilizing effect of measurement in the case of coupling between
two quasiparticles, one inside and another one outside the interference loop, as well as in the case
of bulk-edge coupling. We demonstrate the function of the Zeno effect by analogy with a damped harmonic oscillator, and make estimates regarding the relevance of such an effect to non-Abelian interferometry experiments.

The second effect, described in Section~\ref{polarized}, is closely related to the absorbtion of bulk quasiparticles by the edge as described in \cite{Rosenow08a,Bishara09b,Rosenow09a}. Here, we examine the case in which two quasiparticles near the same edge are both coupled to it with
some strength. We find that the chiral nature of the edge and the Majorana
character of both the edge current and the bulk quasiparticles leads to a degree
of polarization for the coupled quasiparticles even when they are not completely absorbed. We calculate the exact polarization as a function of the coupling strengths and temperature in the limit of a fast edge mode. In particular, we find that the polarization is greatest when the coupling strengths are comparable or when there is an energy splitting between the qubit states.
%

While the parameters involved in the experiments of Willett et
al.\cite{Willett09a,Willett10a} lead us to the conclusion that an appreciable
Zeno effect is unlikely, in Section~\ref{numbers} we estimate that a typical
polarization induced by coupling of the nearest two quasiparticles to the edge
may be around $16\%$, which in turn may explain the presence of an $e/4$ signal
in these experiments despite edge-bulk interactions that would otherwise destroy it.

\section{Ising anyons} \label{Ising}

The edge excitations of the Moore--Read FQH state at $\nu=5/2$ consist of chiral bosonic excitations carrying charge $e/2$ and neutral chiral fermionic modes \cite{Milovanovic96}. The non-Abelian quasiparticles carry charge $e/4$ and correspond to a twist in the boundary conditions for fermions. In the bulk these quasiparticles may be thought of as charged vortices with bound Majorana (i.e. real fermionic) zero modes. Ignoring the charge sector, these quasiparticles can can associated with (the chiral parts of) the fields $I$, $\psi$ and $\sigma$ appearing in the conformal field theory of the critical Ising model. The non-Abelian nature of the $\sigma-$particles is manifest in the fusion rules for the Ising spin field: $\sigma \times \sigma = I + \psi$. This translates into the the statement that the combined state of two $\sigma-$particles may or may not contain a fermion.

The edge of a $p+ip$ superconductor (or, equivalently, a topological superconductor induced by a proximity effect on the surface of a topological
insulator or inside a semiconductor with the strong spin-orbit coupling) lacks charge modes yet its chiral excitations can be associated with the same Ising fields, $I$, $\psi$ and $\sigma$ \cite{Fendley07a}. The non-Abelian anyons in this setting are unpaired Majorana modes bound to vortices \cite{Volovik99,Read00}. The complex fermion mode associated with two such excitations may be either occupied or unoccupied; these two states can span the Hilbert space of a topological qubit. Clearly, the state of such a qubit is flipped whenever a fermionic $\psi-$mode tunnels between the qubit and its surroundings.

For the purpose of this paper we will not distinguish between the cases with charged bosonic edge modes (FQHE) or only neutral excitations ($p+ip$ SC or its equivalents). In both cases we will loosely refer to the quasiparticles as Ising anyons despite the fact that they may differ from Ising by Abelian factors, e.g. can be obtained from Ising fields through products or cosets with U$(1)$ sectors. Importantly, in all these cases the fermionic modes remain neutral and hence their tunnelling cannot be inhibited by Coulomb energy considerations.

A real chiral fermionic edge mode is described by the Lagrangian
\begin{equation}
\label{edge-lagrangian}
L_{\mathrm{edge}} = \frac{\rmi}{2}\int\!\mathrm{d}x \,\psi(\partial_t+v\partial_x )\psi,
\end{equation}
where the normaliazion has been chosen to make the fields
obey the anticommutation relation
\begin{equation}\label{anticommutator}
\{\psi(x),\psi(y)\}=2\delta(x-y).
\end{equation}
With this normalization,
\begin{equation}
\label{correlator}
\langle{\psi(x)\psi(y)}\rangle \sim \frac{1}{\rmi\pi}\,\frac{1}{x-y}
\end{equation}

We shall describe the localized Majorana bound states in the bulk of the system via operators $\gamma_i$ such that $\gamma_i^\dagger=\gamma_i$ and
\begin{equation}
\{\gamma_i,\gamma_j\}=2\delta_{ij}.
\end{equation}
These Majorana bound states act like Ising $\sigma$ fields, in that we may combine two of the bound-state operators to make a normal fermionic degree of freedom, as $\hat{f}=(\gamma_1+\rmi\gamma_2)/2$, so $\hat{f}^\dagger\hat{f}=(1+\rmi\gamma_1\gamma_2)/2$ has eigenvalues $0$ and $1$. In this basis, $\rmi\gamma_1\gamma_2$ has eigenvalues of $\pm1$ and acts as a Pauli matrix $\bsigma_{\!z}\equiv \rmi\gamma_1\gamma_2$.

\section{Zeno effect}
\label{zeno}

Consider a system in which a qubit consisting of two Majorana bound states
$\gamma_1$ and $\gamma_2$ accumulates error via interaction with a separate
Majorana mode. The qubit can take two states corresponding to the eigenvalues of
$\bsigma_{\!z}=\rmi\gamma_1\gamma_2$. We note two distinct forms of the error
Hamiltonian relevant to measurements in systems with a chiral Majorana edge.
First, the qubit system may interact with a localized defect outside the interferometric loop. Second, the qubit
may interact with an edge. These two cases show qualitatively different
behaviour.

In either case the error Hamiltonian is of the form
$H_{\mathrm{err}}=\rmi\lambda\gamma_2\xi$, where $\xi$ is a Majorana operator that
may either be associated with a Majorana bound state or with a Majorana edge.
The $\bsigma_{\!z}$ eigenvalue of the qubit shall be continuously measured either by
environmental interactions or by an experimental apparatus in which the result
of the measurement is a priori unknown. We therefore shall average over that
result in our simple description.

The density matrix of the system is then governed by the equation
\begin{equation}\label{master_eq}
\dot{\brho}(t)=-\rmi\,[H_\mathrm{err},\brho(t)]
-\frac{\kappa}{4}\left[\bsigma_{\!z},\left[\bsigma_{\!z},\brho(t)\right]\right],
\end{equation}
where $\kappa$ characterizes the strength of the measurement.
It shall be convenient for our analysis to consider an auxiliary variable
$\bar{\brho}=\rme^{\kappa t}\brho(t)$, in
terms of which Eq.~\ref{master_eq} becomes:
\begin{equation}\label{master_eq-2}
\dot{\bar{\brho}}(t)=-\rmi\left[H_\mathrm{err},\bar{\brho(t)}\right]
-\frac{\kappa}{4}\left[\bsigma_{\!z},\left[\bsigma_{\!z},\brho(t)\right]\right]+\kappa\bar{
\brho}(t).
\end{equation}
If we wish to know the qubit polarization $z=\mathrm{Tr}\big(\bsigma_{\!z}\brho(t)\big)$, we can
note that
\begin{equation}
\dot{z}=\rme^{-\kappa
t}\mathrm{Tr}\big(\bsigma_{\!z}\left(\dot{\bar{\brho}}-\kappa\bar{\brho}\right)\big)
\end{equation}
in order to find
\begin{eqnarray}\label{master eq-3}
\dot{z}(t)&=&\rme^{-\kappa
t}\mathrm{Tr}\big(-\rmi\left[\bsigma_{\!z},H_\mathrm{err}(t)\right]\bar{
\brho}(t)\big)\nonumber\\
&=&-2\int_0^t\mathrm{d}t'\rme^{-\kappa
(t-t')}\mathrm{Tr}\big(\left\{H_\mathrm{err}(t),H_\mathrm{err}(t')\right\}
\bsigma_{\!z}\brho(t')\big)
\nonumber\\&&+\rmi\rme^{-\kappa
t}\mathrm{Tr}\big([\bsigma_{\!z},H_\mathrm{err}(t)]\brho(0)\big),
\end{eqnarray}
where we have used the cyclic nature of the trace and the fact that
$\{\bsigma_{\!z},H_\mathrm{err}\}=0$.

We consider two forms for the extra Majorana mode interacting with the qubit. In
the first case, $\xi=\gamma_3$ is simply a third localized Majorana state. In
this case, we can map $i\gamma_2\xi\rightarrow\bsigma_{\!y}$ to find
\begin{eqnarray}
\label{case1}
\dot{z}(t)=-4\lambda^2\int_0^t\mathrm{d}t'\rme^{-\kappa (t-t')}z(t')
+2\lambda \rme^{-\kappa t}\mathrm{Tr}\big(\bsigma_{\!x}\brho(0)\big)\nonumber\\
0=\ddot{z}+\kappa\dot{z}+4\lambda^2z.
\end{eqnarray}
This simple harmonic oscillator equation shows the basic function of the Zeno
effect. The measurement strength acts as a drag term on the oscillator, so that
as $\kappa\rightarrow\infty$, the quantum dynamics of the system is frozen (i.e.
$\dot{z}=0$).

On the other hand, for a qubit coupled to an edge 
$\xi=l_0^{1/2}\int\mathrm{d}xf(x)\psi(x-vt)$, where $l_0$ is the short range
cutoff of the theory (the magnetic length, in the quantum Hall setting) and where the edge Majorana mode $\psi$ is governed by the
Lagrangian given in \eref{edge-lagrangian}
and obeys the anticommutation relation $\{\psi(x),\psi(y)\}=2\delta(x-y)$.
We have then that
\begin{equation}
\label{case2}
\dot{z}(t)=-4\lambda^2 l_0\int_0^t\mathrm{d}t'\rme^{-\kappa
(t-t')}\int\mathrm{d}xf(x)f(x-v(t-t'))z(t'),
\end{equation}
where we assume that the edge is in thermal equilibrium at $t=0$, so
$\mathrm{Tr}\big([\bsigma_{\!z},H_\mathrm{err}(t)]\brho(0)\big)
\propto\langle\psi(x-vt)\rangle=0$.

Importantly, if the qubit interacts with only a single point on the edge (i.e.
$f(x)=\delta(x)$) then the above equation reduces to
\begin{equation}
\dot{z}=-2\frac{\lambda^2l_0}{v}z
\end{equation}
There is no Zeno effect here due to the measurement of the qubit. Instead of
oscillating, the qubit undergoes a simple decay, which does not allow the
measurement time to affect the qubit.

We can relate these two cases by including a finite interaction range in the error Hamiltonian $H_\mathrm{err}$. A particularly simple choice for the form factor of the interaction is \mbox{$f(x)=({\pi a})^{-1} K_0\left({\abs{x}}/{a}\right)$}.\footnote{ This may be thought of as a tractable approximation to the more realistic $f(x)\propto \rme^{-\sqrt{(d^2+x^2)/a^2}}$, where $d$ is the distance from the Majorana bound state to the edge.} In this case, we have
\begin{eqnarray}
\dot{z}(t)=-\frac{2\lambda^2 l_0}{a}\int_0^t\mathrm{d}t'\rme^{-(\kappa+v/a)
(t-t')}z(t')\nonumber\\
0=\ddot{z}+\left(\kappa+\frac{v}{a}\right)\dot{z}+\frac{2\lambda^2 l_0}{a}z.
\end{eqnarray}
Note that this reduces to the previous case \eref{case1} when the edge mode
velocity tends to zero.

Using the above equation, however, we can see that the Zeno effect is likely to
have limited influence in interferometric experiments of Willett et al.\ \cite{Willett09a,Willett10a}. Using parameters relevant
to these experiments, the Zeno parameter $\kappa$ is at most $\delta
I^2/8S\sim\delta I/4e$, where $\delta I=I_0 \Delta R_{xx}/R_{xy}$, is the
portion of the signal current coming from the tunnelling of $e/4$ quasiparticles
and $S$ is the spectral density of this current. With $\Delta R_{xx}=2\Omega$,
$I_0=2$nA and $R_{xy}=2h/5e^2$, this is only $604$kHz, whereas the decay from
edge motion is around $v/a\gtrsim v/L\sim22$GHz at the least, given a side length of
$L=0.45\mu$m for the interferometer and a neutral edge velocity of $v=10000$m/s.
The Zeno effect is thus insignificant in the Willett et al. experiment, and
would be for any such $nu=5/2$ FQHE interferometer unless the size of the signal
current is significantly increased.

\section{Edge induced polarization}\label{polarized}
We now expand our analysis to include two Majorana bound states $\gamma_1$ and
$\gamma_2$ forming a qubit that both interact with the edge at different points $x_1$
and $x_2$. For simplicity, we shall assume a $\delta$-like interaction with the
edge. The interaction picture Hamiltonian is given by
\begin{equation}\label{ham}
H(t)=-\rmi\lambda_1\gamma_1\psi(x_1-vt)-\rmi\lambda_2\gamma_2\psi(x_2-vt).
\end{equation}
Note that we have absorbed the short range cutoff $l_0$ into the definitions of $\lambda_1$ and $\lambda_2$, which now have units of ${\sqrt{\rm{length}}}/{\rm{time}}$.
For compactness of notation, we define $\psi_i(t)=\psi(x_i-vt)$.
\Eref{master eq-3} then becomes
\begin{eqnarray}\label{master eq-4}\fl
\dot{z}&=&-2\,\frac{\lambda_1^2
+\lambda_2^2}{v}\,z
\nonumber
\\
\fl
&{ }&+2\rmi\lambda_1\lambda_2
\int_0^t\mathrm{d}t_1 \rme^{-\kappa(t-t_1)}
\mathrm{Tr}\Big(\big(\left[\psi_1(t),\psi_2(t_1)\right]-\left[\psi_2(t),
\psi_1(t_1)\right]\big)\brho(t_1)\Big).
\end{eqnarray}
Because an interferometric measurement of this system is made using $\sigma$
quasiparticles moving at the same neutral velocity $v$ as the edge Majoranas,
the actual expectation value that appears in the result of such a measurement is not
$z$ but rather
\begin{equation}
\tilde{z}(t)=\langle \rmi\gamma_1(t)\gamma_2(t+\Delta x/v)\rangle=\mathrm{Tr}\big(
\rmi\gamma_1(0)\gamma_2(\Delta x/v)\brho(t)\big),
\end{equation}
where $\gamma_i(t)=U^\dagger(t)\gamma_i U(t)$ and
$U(t)=T_\leftarrow\exp\left(-\rmi\int_0^t\mathrm{d}t'H(t')\right)$. ($T_\leftarrow$
indicates time ordering with time increasing to the left).
However, the time scale $\Delta x/v$ is generally significantly shorter than other
experimentally relevant timescales. For simplicity in our estimates of the
polarization effect, we take $\Delta x/v\rightarrow 0$ in our final result. This fast-edge approximation also allows us to disregard the fermion parity of the edge inside the interferometric loop (as treated, e.g., in \cite{Rosenow09a}), as fermions leaving the qubit interaction area will also immediately leave the area of the interferometer.

Repeated application of equations~\eref{anticommutator} and \eref{master_eq}, along with the fact that $x_2>x_1$,
allows us to set up recursion relations for the trace terms of the type in Eq.~(\ref{master eq-4}). For example,
\begin{eqnarray}
\fl
\int_0^t\mathrm{d}t_1 \rme^{-\kappa(t-t_1)}h(t-t_1)\mathrm{Tr}\Big(\left[\psi_1(t),\psi_1(t_1)\right]\brho_I(t_1)\Big)=
\nonumber\\
\int_0^t\mathrm{d}t_1 \rme^{-\kappa(t-t_1)}f(t-t_1)\langle\left[\psi_1(t),\psi_1(t_1)\right]\rangle,
\end{eqnarray}
where $h(t-t_1)=f(t-t_1)+\frac{2\lambda_1^2}{v}\int_{t_1}^tf(t-t')$. (For a full derivation of this type, see the Appendix.)
Solving the resulting integral equations leads to the following expression for $z$:
\begin{eqnarray}
\fl
\dot{z}=&-& 2
\frac{\lambda_1^2+\lambda_2^2}{v}z+2\rmi\lambda_1\lambda_2\left\{\int_{0}^{t}\mathrm{d}u \, \rme^{-\kappa u} \left[\rme^{-\frac{2\lambda_2^2 u}{v}}\,g\!\left(u+\frac{\Delta x}{v}\right)
- \rme^{-\frac{2\lambda_1^2 u}{v}}\,g\!\left(u-\frac{\Delta x}{v}\right)\right]
\right.
\nonumber\\
\fl
&+&\Theta\!\left(t-\frac{\Delta x}{v}\right)
\frac{2\lambda_1^2}{\lambda_2^2-\lambda_1^2}
\int_{0}^{t-\Delta x/v}\mathrm{d}u\left[
\rme^{-\left(\kappa+\frac{2\lambda_2^2}{v}\right)u}
-\rme^{-\left(\kappa+\frac{2\lambda_1^2}{v}\right)u}\right]
g\!\left(u+\frac{\Delta x}{v}\right)
\nonumber\\
\fl
&-&\left.\Theta\!\left(t-\frac{\Delta x}{v}\right)\frac{4\lambda_1^2}{v}
\int_0^{\Delta x/v}\mathrm{d}u_1\int_{\Delta
x/v}^t\mathrm{d}u_2
\rme^{-\left(\kappa+\frac{2\lambda_1^2}{v}\right)
\left(u_1+u_2-\frac{\Delta x}{v}\right)} g\left(u_2-u_1\right)\right\}
\end{eqnarray}
where $g(u)=\langle\left[\psi(-v u),\psi(0)\right]\rangle$ and $\Theta(x)$ is a Heaviside step function.
As $t\rightarrow\infty$, the polarization $z$ reaches an equilibrium value of
\begin{eqnarray}
z_\infty&=&\frac{\rmi\lambda_1\lambda_2v}{\lambda_2^2-\lambda_1^2}
\int_0^\infty\mathrm{d}u\left(\rme^{-(\kappa+2\lambda_2^2/v)u}
-\rme^{-(\kappa+2\lambda_1^2/v)u}\right)g(u+\Delta
x/v)\nonumber\\&+&\frac{\kappa}{(\kappa+2\lambda_1^2/v)}\frac{\rmi\lambda_1\lambda_2v
}{\lambda_1^2+\lambda_2^2}\int_{-\infty}^\infty\mathrm{d}u\,
\rme^{-(\kappa+2\lambda_1^2/v)\abs{u}}g(u+\Delta x/v),
\end{eqnarray}
where we have used that $g(-u)=-g(u)$.
In the experimentally relevant limit $\kappa\ll 2\lambda_2^2/v$, we have
\begin{equation}
z_\infty=\frac{i\lambda_1\lambda_2v}{\lambda_2^2-\lambda_1^2}
\int_0^\infty\mathrm{d}u\left(\rme^{-2\lambda_2^2u/v}
-\rme^{-2\lambda_1^2u/v}\right)g(u+\Delta x/v).
\end{equation}
In the limit of small $\Delta x/v$ we may take $g(u)=2\rmi/\pi v u$ to find:\footnote{A result of a similar form was obtained by Rosenow et al.~\cite{Rosenow08a} for bulk Majoranas coupled to opposite edges of the interferometer.}
\begin{equation}
\label{eq-0temp}
z_\infty=-\frac{2\lambda_1\lambda_2}{
\pi(\lambda_2^2-\lambda_1^2)}\ln\frac{\lambda_1^2}{\lambda_2^2}
\end{equation}
This result is dependent only on the ratio $\lambda_1/\lambda_2$ of the coupling constants, rather than their magnitude. However, the relaxation time for the system is $(2\lambda_1^2/v+2\lambda^2_2/v)^{-1}$, so Majorana bound states that are only weakly coupled to the edge would require a long time to reach this equilibrium.
It is interesting to note that the highest polarization allowed for two Majorana bound states that interact only through the edge in this way is $2/\pi$. This happens to be the correlation between two adjacent Majorana fermions on the edge in the discrete model of this system put forward by Rosenow et al.~\cite{Rosenow09a}.

In order to obtain a higher degree of polarization, we must allow for an energy splitting between the two possible qubit states. A Hamiltonian term of the form $H_\epsilon={\epsilon}\bsigma_{\!z}/2=\rmi{\epsilon}\gamma_1\gamma_2/2$ may be induced either by interaction between the bound states $\gamma_1$ and $\gamma_2$ (see e.g.~\cite{Baraban09,Cheng09,Bonderson09b}) or by a non-locally generated energy splitting do to the circulating current of $\mathbf{\sigma}-$quasiparticles around the edge of the interferometer, as in \cite{Bonderson10b,Clarke10a}. Adding this term to the Hamiltonian (\ref{ham}) and repeating the above analysis (leaving $\kappa=0$ for simplicity and taking $\Delta x/v\rightarrow 0$ as before), we obtain the more general form
\begin{equation}
\label{eq-eta0temp}
z_\infty=-\frac{2\eta}{\pi\sqrt{\eta^2-1}}\arctan\sqrt{\eta^2-1},
\end{equation}
where $\eta=\left({v\epsilon+2\lambda_1\lambda_2}\right)/
\left({\lambda_1^2+\lambda_2^2}\right)$. This reduces to \eref{eq-0temp} for $\epsilon=0$.

\section{Possible experimental relevance}
\label{numbers}
\begin{figure}[htb]
\begin{center}
\includegraphics[width=0.8\columnwidth]{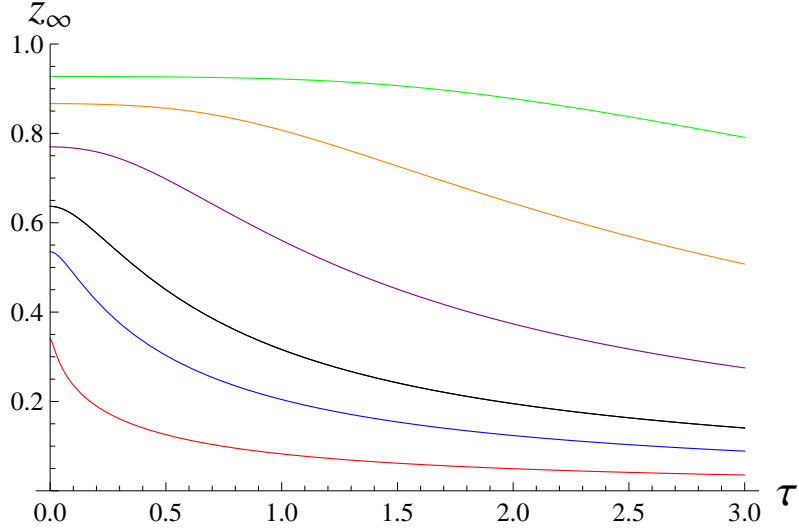}
\caption{The equilibrium polarization due to tunnelling between two bulk
Majorana bound states and the edge is plotted against the reduced temperature
$\tau=\pi vT/(\lambda_1^2+\lambda_2^2)$, where $kT_i=2\hbar\lambda_i^2/v$, and $\lambda_1$ is the stronger
bulk-edge coupling. The series of lines represent different values of $|\eta|$: 0.25 (lowest polarization, red curve), 0.63 (estimated experimental value, blue curve), 1.0 (largest attainable polarization without direct coupling between the bound states, black curve), 2.0, 4.0, and 8.0 (highest, green curve).
The polarization saturates at $T=0$ for $\eta=\mp 1$ at a value of $\pm 2/\pi$, and for $\eta=\mp\infty$ at a value of $\pm 1$. (Colour online).}
\label{fig:polarization}
\end{center}
\end{figure}
In the experiments \cite{Willett09a,Willett10a} of Willett et al., the $\nu=5/2$ plateau is found at $6.5$~Tesla,
corresponding to a magnetic length $l_0=\sqrt{\hbar/(eB)}\sim 10$nm. The area of
the interference loop is altered by changing a side gate voltage. Because the
relevant interference occurs between edge modes with charge $e/4$, the areal
change corresponding to one oscillation is $4h/eB$, whereas the length of the
edge containing the side gate is $0.45\mu$m. There are approximately six
oscillations in each run of $e/4$ periods, corresponding to a distance between
quasiparticles of approximately $d\approx 34$nm. From the numerical calculations of
Baraban et al.~\cite{Baraban09}, we expect the energy scale for tunnelling at
this distance to be $E_1\sim\exp\left[{-(d-l_0)/2.3 l_0}\right]$K$\approx350$mK. The neutral edge
velocity is of order $10^4$ m/s, so this corresponds to a decay rate of roughly
$2\lambda_1^2/v=2E_1^2l_0/v\sim4$GHz for the closest quasiparticle to
the edge and $2\lambda_2^2/v=2E_2^2l_0/v\sim0.5$GHz for the second
closest quasiparticle, which we expect to be at most around $70$nm from the
edge.

Note that since $v/\Delta x\gtrsim v/L\sim 22$GHz, as calculated in Sec.~\ref{zeno}, these numbers justify the approximation $2\Delta x\lambda_i^2/v^2\rightarrow 0$ that we have used above.

We expect then that $\lambda_1^2/\lambda_2^2\sim 8$ and take $\epsilon=0$, in which case the
\eref{eq-0temp} leads to a polarization of approximately $52\%$ at $T=0$.
However, the experiment of Willett et al. is conducted at a temperature of
$T=25$mK. For a finite temperature, we must replace the edge correlation
function $g(u)=2\rmi/\pi v u$ with $g(u,T)=2\rmi T/\left[v\sinh(\pi T u)\right]$. The resulting
polarization has the form $z=z(\eta,\tau)$ where $\eta=(v\epsilon+2\lambda_1\lambda_2)/(\lambda_1^2+\lambda_2^2)$ as before and $\tau=\pi vT/(\lambda_1^2+\lambda_2^2)$:
\begin{equation}\label{eq-etafintemp}
\fl
z_\infty(\eta,\tau)= -\frac{i\eta}{\pi\sqrt{\eta^2-1}} \left[ \Psi\left(\frac{\pi\tau+1-i\sqrt{\eta^2-1}}{2\pi\tau}\right)
-\Psi\left(\frac{\pi\tau+1+i\sqrt{\eta^2-1}}{2\pi\tau}\right)\right],
\end{equation}
where $\Psi(x)$ is the digamma function. (See Appendix for technical details leading to this result.) In \fref{fig:polarization}, $z_\infty$ is plotted as a function of the reduced temperature $\tau$ for various values of $\eta$.
For the experimental values discussed above, at a temperature of $T=25$mK, we have $\abs{\eta}\approx0.63$ and $\tau\approx 2.5$. This leads finally to a polarization of $\sim 16\%$.

Interestingly, the degree of polarization can be quite high if $\epsilon \gg \lambda_{1,2}^2/v$ (e.g., for quasiparticles that are close together and far from the edge) although the relaxation time for reaching this equilibrium value is still $\sim v/\lambda^2$. In particular, the polarization can exceed $2/\pi$ provided that the two quasiparticles interact with one another inducing a non-zero energy splitting $\epsilon$. Likewise, for non-zero $\epsilon$, the polarization persists even if only one quasiparticle is coupled directly to the edge.

\section{Discussion}
In this note, we have examined two possible mechanisms for the
presence of a non-Abelian signal in Ising-type anyonic
interferometry despite the coupling of the Majorana bound
states comprising the qubit to the edge. The first of these,
the quantum Zeno effect, is in principle present in any such
system due to the continuous measurement. However, as we have
shown, it does not provide a viable explanation for the experimental data
reported in \cite{Willett09a,Willett10a} due to the small value
of the measurement strength parameter $\kappa$ as compared
with the frequency at which the edge states travel around the
interferometric loop.

The second mechanism, that of qubit polarization due to the
coupling of multiple bulk Majorana bound states to the edge,
is more likely to explain the observed data. This mechanism is
complementary to the results of Rosenow et al.
\cite{Rosenow08a,Rosenow09a} and Bishara and Nayak \cite{Bishara09b}, which describe the effective absorbtion
of bulk Majorana states by the edge as the coupling is
increased. Here, we have demonstrated that even when a pair of
quasiparticles are not fully absorbed their interaction with
the edge can cause a significant polarization of the qubit.
This polarization is maximized when the coupling constants
have equal magnitude and at $T=0$ depends only on their ratio.
We estimate, based on the parameters relevant to
experiments \cite{Willett09a,Willett10a}, that the qubit
polarization may be as high as $16\%$ when the quasiparticles
are coupled to the same edge. This is significantly higher
than the polarization caused by coupling to opposite edges, as
found in \cite{Rosenow08a}.
An extension of this result to the more realistic situation of
many quasiparticles in the bulk shall be the subject of future
research.

\ack
The authors are grateful to A.~Korotkov and S.~Simon for helpful discussions. DC and KS are supported in part by the DARPA-QuEST program. KS is supported in
part by the NSF under grant DMR-0748925.

\appendix
\section*{Appendix}
\setcounter{section}{1}
\label{sec:appendix}
We begin with the Hamiltonian
\begin{equation}\label{apeq-ham}
H(t)=H_\lambda+H_\epsilon=-\rmi\lambda_1\gamma_1\psi_1(t)
-\rmi\lambda_2\gamma_2\psi_2(t)+ \frac{\epsilon}{2}\bsigma_{\!z},
\end{equation}
where $\bsigma_{\!z}=\rmi\gamma_1\gamma_2$ and $\psi_i(t)=\psi(x_i-vt)$.
For simplicity, we assume the measurement parameter, $\kappa$ may be safely set to $0$.
We set $\trho=\rme^{\rmi \epsilon\bsigma_{\!z} t/2}\brho(t)\rme^{-\rmi \epsilon\bsigma_{\!z} t/2}$, and \mbox{$\tH=\rme^{\rmi \epsilon\bsigma_{\!z} t/2}H_\lambda(t)\rme^{-\rmi \epsilon\bsigma_{\!z} t/2}=H_\lambda(t)\rme^{-\rmi \epsilon\bsigma_{\!z} t}$}.
Then
\begin{equation}
\dot{\trho}(t)=-\rmi\left[\tH(t),\trho(t)\right]
\end{equation}
so
\begin{eqnarray}\label{apeq-z}
\fl
\dot{z}(t)&=&\Tr\left(\bsigma_{\!z}\brho\right)=-2\int_0^t\mathrm{d}t_1\Tr\left(\left\{\tH(t),\tH(t_1)\right\}\bsigma_{\!z}\trho(t_1)\right)
\nonumber\\
\fl
&=&-2\int_0^t\mathrm{d}t_1\Tr\left(\left\{H_\lambda(t),H_\lambda(t_1)\right\}\bsigma_{\!z}\trho(t_1)\right)
\cos \epsilon(t-t_1)
\nonumber\\
\fl
&&-2\,\rmi\int_0^t\mathrm{d}t_1\Tr\left(\left[H_\lambda(t),H_\lambda(t_1)\right]\trho(t_1)\right)
\sin \epsilon(t-t_1)
\nonumber\\
\fl
&=&-2\,\frac{\lambda_1^2
+\lambda_2^2}{v}\,z(t)-\Theta\left(t-\frac{\Delta x}{v}\right)\frac{4\lambda_1\lambda_2}{v}\,\sin\!\frac{\epsilon\Delta x}{v}\,z\!\left(t-\frac{\Delta x}{v}\right)
\nonumber\\
\fl
&&+2\,\rmi\lambda_1\lambda_2
\int_0^t\mathrm{d}t_1
\Tr\left(\Big(\left[\psi_1(t),\psi_2(t_1)\right]-\left[\psi_2(t),
\psi_1(t_1)\right]\Big)\trho(t_1)\right)\cos \epsilon(t-t_1).
\nonumber\\
\fl
&&-2\,\rmi
\int_0^t\mathrm{d}t_1
\Tr\left(\Big(\lambda_1^2\left[\psi_1(t),\psi_1(t_1)\right]+\lambda_2^2\left[\psi_2(t),
\psi_2(t_1)\right]\Big)\trho(t_1)\right)\sin \epsilon(t-t_1),
\nonumber\\
\fl
\end{eqnarray}
where we have used that $\{\psi(x),\psi(y)\}=2\delta(x-y)$ and the fact that $x_2>x_1$.

We now note that for any function $f(t)$, we have that
\begin{eqnarray}\label{apeq-f}
\fl
\int_0^t\mathrm{d}t_1f(t-t_1)\Tr\left(\left[\psi_i(t),\psi_j(t_1)\right]\trho(t_1)\right)=
\int_0^t\mathrm{d}t_1f(t-t_1)\langle\left[\psi_i(t),\psi_j(t_1)\right]\rangle
\nonumber\\
-\int_0^t\mathrm{d}t_1\int_0^{t_1}\mathrm{d}t_2\int_0^{t_2}\mathrm{d}t_3f(t-t_1)\Tr
\big(X_{ij}(t,t_1,t_2,t_3)\trho(t_3)\big),
\end{eqnarray}
where
\begin{eqnarray}
X_{ij}&=&\left[\left[\left[\psi_i(t),\psi_j(t_1)\right],\tH(t_2)\right],\tH(t_3)\right]
\nonumber\\
&=&\sum_{k,l}\lambda_k\lambda_l \Big(\left\{\left[\left[\psi_i(t),\psi_j(t_1)\right],\psi_k(t_2)\right],\psi_l(t_3)\right\}
\nonumber\\
&{ }& \qquad\qquad \times \big(\delta_{kl}\sin \epsilon(t_2-t_3)- \varepsilon_{kl}\cos \epsilon(t_2-t_3)\big)\, \rmi\bsigma_{\!z}
\nonumber\\
&{ }&+ \left[\left[\left[\psi_i(t),\psi_j(t_1)\right],\psi_k(t_2)\right],\psi_l(t_3)\right]
\nonumber\\
&{ }& \qquad\qquad \times
\big(\delta_{kl}\cos \epsilon(t_2-t_3)+\varepsilon_{kl}\sin \epsilon(t_2-t_3)\big)\Big),
\end{eqnarray}
where we have used that $\{\gamma_k,\gamma_l\}=2\delta_{kl}$ and  $[\gamma_k,\gamma_l]=-2\rmi\varepsilon_{kl}\bsigma_{\!z}$, where $\varepsilon_{kl}$ is the rank 2 antisymmetric tensor.

We may now use the anticommutation relation \mbox{$\{\psi_i(t),\psi_j(t')\}=2\delta\big(x_i\!-\!x_j\!-\!v(t\!-\!t')\big)$} to note that
\begin{eqnarray}
\fl
\left\{\left[\left[\psi_i(t),\psi_j(t_1)\right],\psi_k(t_2)\right],\psi_l(t_3)\right\}&=&
8\delta\big(x_i\!-\!x_l\!-\!v(t\!-\!t_3)\big)
\delta\big(x_j\!-\!x_k\!-\!v(t_1\!-\!t_2)\big)
\nonumber\\
\fl
&-&8\delta\big(x_i\!-\!x_k\!-\!v(t_t\!-\!t_2)\big)
\delta\big(x_j\!-\!x_l\!-\!v(t_1\!-\!t_3)\big)
\end{eqnarray}
and
\begin{eqnarray}
\fl
\left[\left[\left[\psi_i(t),\psi_j(t_1)\right],\psi_k(t_2)\right],\psi_l(t_3)\right]
\nonumber\\
\fl
\quad = 4\left[\psi_i(t),\psi_l(t_3)\right]\delta\left(x_j\!-\!x_k\!-\!v(t_1\!-\!t_2)\right)-
4\left[\psi_j(t),\psi_l(t_3)\right]\delta\left(x_i\!-\!x_k\!-\!v(t\!-\!t_2)\right)
\end{eqnarray}
Inserting these identities into \eref{apeq-f} and using the fact that $x_2>x_1$ leads to the general formula
\begin{equation}
\fl
\sum_{ij}\int_0^t\mathrm{d}t'h_{ij}(t\!-\!t')
{\Tr}\big(\left[\psi_i(t),\psi_j(t')\right]\trho(t')\big)
=\sum_{ij}\int_0^t\mathrm{d}t'f_{ij}(t\!-\!t')\langle\left[\psi_i(t),\psi_j(t')\right]\rangle
\end{equation}
in the limit as $(x_2-x_1)/v\rightarrow 0$, where for general functions $f_{ij}$ we have
\begin{eqnarray}
\label{apeq-g1}
\fl
h_{i1}(t\!-\!t')=f_{i1}(t\!-\!t')+\int_{t'}^t\mathrm{d}t_1 &{}& \left\{\frac{2\lambda_1^2}{v}
\big(f_{i1}(t\!-\!t_1)+2f_{i2}(t\!-\!t_1)\big)\cos \epsilon(t_1\!-\!t') \right.
\nonumber\\
\fl
&{}& \qquad
\left. -  \frac{2\lambda_1\lambda_2}{v}f_{i2}(t\!-\!t_1)\sin \epsilon(t_1\!-\!t')\right\}
\end{eqnarray}
and
\begin{eqnarray}
\label{apeq-g2}
\fl
h_{i2}(t\!-\!t')=f_{i2}(t\!-\!t')+\int_{t'}^t\mathrm{d}t_1 &{}&
\left\{\frac{2\lambda_1\lambda_2}{v}
\big(f_{i1}(t\!-\!t_1)+2f_{i2}(t\!-\!t_1)\big)\sin \epsilon(t_1\!-\!t') \right.
\nonumber\\
\fl
&{}& \qquad
\left. +\frac{2\lambda_2^2}{v}f_{i2}(t\!-\!t_1)\cos \epsilon(t_1\!-\!t')\right\}.
\end{eqnarray}
We may now make the connection with the original equation \eref{apeq-z} for $z$ in the limit \mbox{$t\rightarrow\infty$, $(x_2-x_1)/v\rightarrow 0$} to find
\begin{equation}\label{apeq-zinf}
z_\infty=\frac{\rmi v}{\lambda_1^2+\lambda_2^2}\sum_{ij}
\int_0^\infty\mathrm{d}t'f_{ij}(t')\langle\left[\psi_i(t'),\psi_j(0)\right]\rangle.
\end{equation}
where we now have fixed $f_{ij}$ according to equations \eref{apeq-g1} and~ \eref{apeq-g2} with
\begin{eqnarray}
h_{11}(t\!-\!t')&=&-\lambda_1^2 \sin \epsilon(t-t')\nonumber\\
h_{12}(t\!-\!t')&=&\lambda_1\lambda_2 \cos \epsilon(t-t')\nonumber\\
h_{21}(t\!-\!t')&=&-\lambda_1\lambda_2 \cos \epsilon(t-t')\nonumber\\
h_{22}(t\!-\!t')&=&-\lambda_2^2 \sin \epsilon(t-t')
\end{eqnarray}
In the limit $(x_2-x_1)/v\rightarrow 0$, $\langle\left[\psi_i(t),\psi_j(0)\right]\rangle=2\rmi T/\left[v\sinh(\pi T t)\right]$ independent of $i$ and $j$. Hence, we need only the function $F(t)=\sum_{ij}f_{ij}(t)/\left(\lambda_1^2+\lambda_2^2\right)$, which obeys the fourth order differential equation
\begin{equation}
\left(\partial_t^2+\epsilon^2\right)
\left[\partial_t^2+\frac{2(\lambda_1^2+\lambda_2^2)}{v}\partial_t+
\left(\epsilon+\frac{2\lambda_1\lambda_2}{v}\right)^2\right]F=0
\end{equation}
with
\begin{eqnarray}
F(0)&=&0\nonumber\\
\partial_t F(0)&=&-\left(\epsilon+\frac{2\lambda_1\lambda_2}{v}\right)\nonumber\\
\partial_t^2F(0)&=& \frac{2(\lambda_1^2+\lambda_2^2)}{v} \left(\epsilon+\frac{2\lambda_1\lambda_2}{v}\right)\nonumber\\
\partial_t^3F(0)&=&\left[\left(\epsilon+\frac{2\lambda_1\lambda_2}{v}\right)^2
-\left(\frac{2(\lambda_1^2+\lambda_2^2)}{v}\right)^2\right]
\left(\epsilon+\frac{2\lambda_1\lambda_2}{v}\right)
\end{eqnarray}
Remarkably, these boundary conditions lead to a relatively simple form for $F$:
\begin{equation}
F(t)=\frac{\rmi\sqrt{-\omega_+\omega_-}}{\omega_--\omega_+}\left(\rme^{\rmi\omega_+ t}-\rme^{\rmi\omega_- t}\right),
\end{equation}
where
\begin{equation}
\omega_\pm=i\frac{\lambda_1^2+\lambda_2}{v}\pm \sqrt{\left(\epsilon+\frac{2\lambda_1\lambda_2}{v}\right)^2-\left(\frac{(\lambda_1^2+\lambda_2^2)}{v}\right)^2}
\end{equation}
Combining this with \eref{apeq-zinf} gives
\begin{equation}
z_\infty=-\int_0^\infty\mathrm{d}t'\frac{2\rmi T}{\sinh(\pi T t)}\frac{\sqrt{-\omega_+\omega_-}}{\omega_--\omega_+}\left(\rme^{\rmi\omega_+ t}-\rme^{\rmi\omega_- t}\right),
\end{equation}
which reduces to \eref{eq-etafintemp} when the integral is performed.
\section*{References}
\providecommand{\newblock}{}


\begin{thebibliography}{10}
\expandafter\ifx\csname url\endcsname\relax
  \def\url#1{{\tt #1}}\fi
\expandafter\ifx\csname urlprefix\endcsname\relax\def\urlprefix{URL }\fi
\providecommand{\eprint}[2][]{\url{#2}}

\bibitem{Moore91}
Moore G and Read N 1991 {\em Nucl. Phys. B\/} {\bf 360} 362--396

\bibitem{Greiter92}
Greiter M, Wen X~G and Wilczek F 1992 {\em Nucl. Phys. B\/} {\bf 374} 567--614

\bibitem{Willett09a}
Willett R~L, Pfeiffer L~N and West K~W 2009 {\em PNAS\/} {\bf 106} 8853--8858

\bibitem{Willett10a}
Willett R~L, Pfeiffer L~N and West K~W 2010 {\em Phys. Rev. B\/} {\bf 82}
  205301 (\textit{Preprint} \eprint{arXiv:0911.0345})

\bibitem{Fu08}
Fu L and Kane C~L 2008 {\em Phys. Rev. Lett.\/} {\bf 100} 096407
  (\textit{Preprint} \eprint{arXiv:0707.1692})

\bibitem{Sau10a}
Sau J~D, Lutchyn R~M, Tewari S and Das~Sarma S 2010 {\em Phys. Rev. Lett.\/}
  {\bf 104} 040502 (\textit{Preprint} \eprint{arXiv:0907.2239})

\bibitem{Alicea10a}
Alicea J 2010 {\em Phys. Rev. B\/} {\bf 81} 125318 (\textit{Preprint}
  \eprint{arXiv:0907.2239})

\bibitem{Lee09}
Lee P~A 2009  (\textit{Preprint} \eprint{arXiv:0907.2681})

\bibitem{Kitaev03}
Kitaev A~Y 2003 {\em Ann. Phys.\/} {\bf 303} 2 (\textit{Preprint}
  \eprint{quant-ph/9707021})

\bibitem{Preskill98}
Preskill J 1998 Fault-tolerant quantum computation {\em Introduction to Quantum
  Computation\/} ed Lo H~K, Popescu S and Spiller T~P (World Scientific)
  (\textit{Preprint} \eprint{quant-ph/9712048})

\bibitem{Freedman03b}
Freedman M~H, Kitaev A, Larsen M~J and Wang Z 2003 {\em Bull. Amer. Math. Soc.
  (N.S.)\/} {\bf 40} 31--38 ISSN 0273-0979 (\textit{Preprint}
  \eprint{quant-ph/0101025})

\bibitem{Fradkin98}
Fradkin E, Nayak C, Tsvelik A and Wilczek F 1998 {\em Nucl. Phys. B\/} {\bf
  516} 704--18 (\textit{Preprint} \eprint{cond-mat/9711087})

\bibitem{DasSarma05}
Das~Sarma S, Freedman M and Nayak C 2005 {\em Phys. Rev. Lett.\/} {\bf 94}
  166802 (\textit{Preprint} \eprint{cond-mat/0412343})

\bibitem{Bonderson06b}
Bonderson P, Shtengel K and Slingerland J~K 2006 {\em Phys. Rev. Lett.\/} {\bf
  97} 016401 (\textit{Preprint} \eprint{cond-mat/0601242})

\bibitem{Bonderson08a}
Bonderson P, Shtengel K and Slingerland J~K 2008 {\em Ann. Phys.\/} {\bf 323}
  2709--2755 (\textit{Preprint} \eprint{arXiv:0707.4206})

\bibitem{Akhmerov09a}
Akhmerov A~R, Nilsson J and Beenakker C~W~J 2009 {\em Phys. Rev. Lett.\/} {\bf
  102} 216404 (\textit{Preprint} \eprint{arXiv:0903.2196})

\bibitem{Fu09b}
Fu L and Kane C~L 2009 {\em Phys. Rev. Lett.\/} {\bf 102} 216403
  (\textit{Preprint} \eprint{arXiv:0903.2427})

\bibitem{Nilsson10a}
Nilsson J and Akhmerov A~R 2010 {\em Phys. Rev. B\/} {\bf 81} 205110
  (\textit{Preprint} \eprint{arXiv:0912.4716})

\bibitem{Sau10c}
Sau J~D, Tewari S and Das~Sarma S 2010  (\textit{Preprint}
  \eprint{arXiv:1004.4702})

\bibitem{Rosenow08a}
Rosenow B, Halperin B~I, Simon S~H and Stern A 2008 {\em Phys. Rev. Lett.\/}
  {\bf 100} 226803 (\textit{Preprint} \eprint{arXiv:0707.4474})

\bibitem{Bishara09b}
Bishara W and Nayak C 2009 {\em Phys. Rev. B\/} {\bf 80} 155304
  (\textit{Preprint} \eprint{arXiv:0906.0327})

\bibitem{Rosenow09a}
Rosenow B, Halperin B~I, Simon S~H and Stern A 2009 {\em Phys. Rev. B\/} {\bf
  80} 155305 (\textit{Preprint} \eprint{arXiv:0906.0310})

\bibitem{Misra77}
Misra B and Sudarshan E~C~G 1977 {\em J. Math. Phys.\/} {\bf 18} 756--763

\bibitem{Milovanovic96}
Milovanovi\'{c} M and Read N 1996 {\em Phys. Rev. B\/} {\bf 53} 13559--13582
  (\textit{Preprint} \eprint{cond-mat/9602113})

\bibitem{Fendley07a}
Fendley P, Fisher M~P~A and Nayak C 2007 {\em Phys. Rev. B\/} {\bf 75} 045317
  (\textit{Preprint} \eprint{cond-mat/0607431})

\bibitem{Volovik99}
Volovik G 1999 {\em JETP Letters\/} {\bf 70}(9) 609--614

\bibitem{Read00}
Read N and Green D 2000 {\em Phys. Rev. B\/} {\bf 61} 10267--10297
  (\textit{Preprint} \eprint{cond-mat/9906453})

\bibitem{Baraban09}
Baraban M, Zikos G, Bonesteel N and Simon S~H 2009 {\em Phys. Rev. Lett.\/}
  {\bf 103} 076801 (\textit{Preprint} \eprint{http://arxiv.org/abs/0901.3502})

\bibitem{Cheng09}
Cheng M, Lutchyn R~M, Galitski V and Sarma S~D 2009 {\em Phys. Rev. Lett.\/}
  {\bf 103} 107001 (\textit{Preprint} \eprint{arXiv:0905.0035})

\bibitem{Bonderson09b}
Bonderson P 2009 {\em Phys. Rev. Lett.\/} {\bf 103} 110403 (\textit{Preprint}
  \eprint{arXiv:0905.2726})

\bibitem{Bonderson10b}
Bonderson P, Clarke D~J, Nayak C and Shtengel K 2010 {\em Phys. Rev. Lett.\/}
  {\bf 104} 180505 (\textit{Preprint} \eprint{arXiv:0911.2691})

\bibitem{Clarke10a}
Clarke D~J and Shtengel K 2010 {\em Phys. Rev. B\/} {\bf 82} 180519
  (\textit{Preprint} \eprint{1009.0302})

\end{thebibliography}
\end{document}